\documentclass[
 reprint,
 superscriptaddress,
 amsmath,amssymb,
 aps,
 prl,
]{revtex4-2}

\usepackage{graphicx}
\usepackage{dcolumn}
\usepackage{bm}
\usepackage{hyperref}
\usepackage{amsmath, amssymb, mathtools, isomath, nccmath}
\usepackage{siunitx}
\usepackage{xspace}
\usepackage{xcolor}
\usepackage[british]{babel}
\usepackage{sidecap}

\newcommand{\Rb}{\ensuremath{^{87}\text{Rb}}\xspace}%
\usepackage{xcolor}

\sidecaptionvpos{figure}{h}

\begin{document}

\title{\bf{Superresolution microscopy of optical fields using tweezer-trapped single atoms}}
\author{Emma Deist}
\affiliation{Department of Physics, University of California, Berkeley, California 94720}
\affiliation{Challenge Institute for Quantum Computation, University of California, Berkeley, California 94720}
\author{Justin A. Gerber}
\affiliation{Department of Physics, University of California, Berkeley, California 94720}
\affiliation{Challenge Institute for Quantum Computation, University of California, Berkeley, California 94720}
\author{Yue-Hui Lu}
\affiliation{Department of Physics, University of California, Berkeley CA 94720}
\affiliation{Challenge Institute for Quantum Computation, University of California, Berkeley, California 94720}
\author{Johannes Zeiher}
\affiliation{Department of Physics, University of California, Berkeley CA 94720}
\affiliation{Max-Planck-Institut f\"{u}r Quantenoptik, 85748 Garching, Germany}
\affiliation{Munich Center for Quantum Science and Technology (MCQST), 80799 Munich, Germany}
\author{Dan M. Stamper-Kurn}
\email[]{dmsk@berkeley.edu}
\affiliation{Department of Physics, University of California, Berkeley, California 94720}
\affiliation{Challenge Institute for Quantum Computation, University of California, Berkeley, California 94720}
\affiliation{Materials Sciences Division, Lawrence Berkeley National Laboratory, Berkeley, California 94720}
\date{\today}

\begin{abstract}

We realize a scanning probe microscope using single trapped $^{87}$Rb atoms to measure optical fields with subwavelength spatial resolution.
Our microscope operates by detecting fluorescence from a single atom driven by near-resonant light and determining the ac Stark shift of an atomic transition from other local optical fields via the change in the fluorescence rate.
We benchmark the microscope by measuring two standing-wave Gaussian modes of a Fabry-P\'{e}rot resonator with optical wavelengths of \SI{1560}{nm} and \SI{781}{nm}.  We attain a spatial resolution of \SI{300}{nm}, which is superresolving compared to the limit set by the \SI{780}{nm} wavelength of the detected light.
Sensitivity to short length scale features is enhanced by adapting the sensor to characterize an optical field via the force it exerts on the atom.
\end{abstract}
\maketitle


Neutral atoms make excellent sensors, owing largely to the identical physical properties of all atoms of a given isotope~\cite{Kitching2011}.
Further, neutral atoms can be isolated from decoherence, enabling highly sensitive measurements of fields, forces, acceleration, rotation, and the passage of time~\cite{budk07nphys,schr14sql,bong19review,ludl15rmp}.
While, to date, atomic sensors have mostly made use of gaseous atomic ensembles, new techniques allow exceptional control of single atoms~\cite{Schlosser2001,Kaufman2012} and of structured arrays of single atoms~\cite{Barredo2016,Endres2016}, motivated by the goals of quantum simulation~\cite{Browaeys2020}, communication~\cite{Reiserer2015}, metrology~\cite{Madjarov2019,Young2020}, and computation~\cite{Weitenberg2011,Levine2019}. 

A number of techniques use cold atoms for spatial tomography of material properties and electromagnetic fields~\cite{Guthorlein2001,veng07mag,Brantut2008,Gierling2011,Lee2014,Yang2017}.
Complementary techniques use electromagnetic fields to perform superresolution microscopy of cold atom systems~\cite{McDonald2019,Subhankar2019}. 
Here, we harness the ability to trap, position, and detect single neutral atoms to construct a scanning probe quantum sensor~\cite{Bian2021} that measures optical fields with high spatial resolution.
The sensing medium is a single $^{87}$Rb atom trapped within a tightly focused optical tweezer trap and driven with near-resonant light.
By measuring the optical fluorescence rate, we determine the shift induced on the atomic resonance frequency by local optical fields and, thereby, the local background optical intensity.  
We apply our sensor to optical test patterns formed by long-wavelength (LW, \SI{1560}{nm}) and short-wavelength (SW, \SI{781}{nm}) TEM$_{00}$ standing-wave modes of a Fabry-P\'{e}rot optical resonator.
We measure a sensor spatial resolution of \SI{300}{nm}, below the resolution limit set by the detected fluorescence light at a wavelength of \SI{780}{nm}, thus achieving superresolution. 


\begin{figure}
    \centering
    \includegraphics{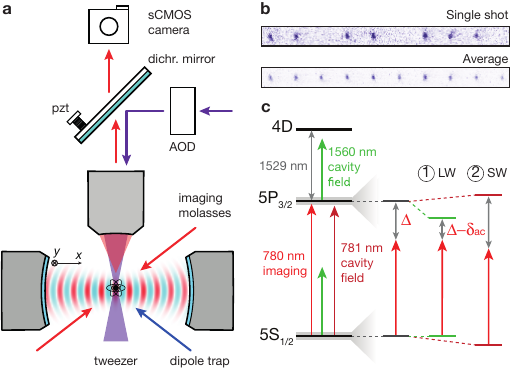} 
    \caption{Overview of the experiment.
    (a) Atoms are loaded from an optical dipole trap into an array of optical tweezers, centered within an optical cavity.  The radial ($y$) position of tweezers along the array is controlled using an acousto-optical deflector (AOD), while the axial ($x$) position is controlled using a piezoactuated mirror.  Atomic fluorescence is imaged through a $\mathrm{NA}=0.5$ objective (used also to focus tweezer light) onto a scientific CMOS (sCMOS) camera.
    (b) Single-shot (top) and averaged (bottom) fluorescence images of ten-atom tweezer array.
    (c) ac Stark shifts are applied to the \Rb $5S_{1/2}$ to $5P_{3/2}$ imaging transition at \SI{780}{nm} by light at wavelengths of \SI{1560}{nm} (LW, green arrow) and \SI{781}{nm} (SW, dark red arrow), both supported by the optical cavity. LW light reduces the imaging transition frequency, due to the dominant downward shift of the excited state caused by the proximity of the $5P_{3/2}$ to $4D$ transitions near \SI{1529}{nm}.
    SW light increases the imaging transition frequency and exerts a strong potential on ground-state atoms.
    }
    \label{fig:Fig_1}
\end{figure}

Operation of the sensor relies on basic properties of light-atom interactions.
An atom in free space scatters light at a rate given by
\begin{equation}
\Gamma_\mathrm{sc} \propto \frac{1}{(\omega_L - \omega_0 - \delta_\mathrm{ac})^2}. \label{eq:brightness}
\end{equation}
Here, we consider the scattering of imaging light at a frequency $\omega_L$ that is near that of a single atomic transition, with the atomic resonance frequency being the sum of $\omega_0$, the bare resonance frequency, and $\delta_\mathrm{ac}$, the transition ac Stark shift. 
We consider the ac Stark shift due to linearly polarized optical fields with detunings much larger than the atomic state hyperfine splittings, and thus treat the atom as a two-level system, ignoring degeneracies and optical polarization effects. 
We require the detuning of the imaging light to be large compared to the atomic resonance linewidth and neglect saturation by considering the weak scattering regime. 
The transition ac Stark shift is determined by the intensity $I$ of a local optical field of frequency $\omega$ as $\delta_{\mathrm{ac}} = - \left(2 \hbar c \epsilon_0\right)^{-1} \left[ \alpha_e(\omega) - \alpha_g(\omega) \right] I$, where $\alpha_{g,e}(\omega)$ are the scalar dynamical electrical polarizabilities of the ground and excited states~\cite{Grimm2000}. 
These polarizabilities are fixed for $^{87}$Rb atoms; thus, the measurement of $\delta_{\mathrm{ac}}$ realizes an absolutely calibrated light intensity meter.

Our sensor employs a one-dimensional array of as many as ten atoms trapped individually in optical tweezer traps.
The tweezers, each formed by focusing light at a wavelength of \SI{808}{nm} through a $\mathrm{NA}=0.5$ objective to a Gaussian beam waist of around \SI{750}{nm}, are located near the center of an in-vacuum, near-concentric Fabry-P\'{e}rot optical cavity, whose mirrors are coated to be highly reflective for LW and SW light. 
The array is oriented perpendicular to the cavity axis and translated along the cavity axis using a piezo-controlled mirror; see Figs.~\ref{fig:Fig_1}(a) and (b).

We load atoms into the tweezers by overlaying the tweezers on a large-volume optical trap containing a gas of $^{87}$Rb atoms at a temperature of \SI{30}{\micro\K}.
The atoms are then exposed both to counterpropagating fluorescence imaging beams, with a variable red detuning $\Delta = \omega_L - \omega_0 < 0$ from the $D_2$ $F=2 \rightarrow F^\prime = 3$ optical transition, and also to repump light resonant with the $D_2$ $F=1 \rightarrow F^\prime = 2$ transition. 
The dipole trap is extinguished after \SI{10}{ms}. 
The imaging light reduces the atom number in each tweezer to either zero or one~\cite{Schlosser2001}, cools the atoms to a thermal energy of roughly a tenth of the depth of the tweezer trap [set in the range $k_B \times (0.25-1.5)$ \SI{}{mK} for different experiments], and generates atomic fluorescence.

\begin{figure}
    \centering
    \includegraphics{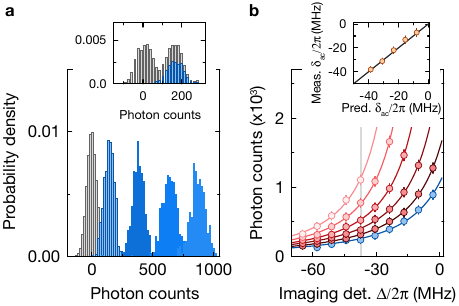}
    \caption{Demonstration of sensor operation.
    (a) Background-subtracted integrated photon counts for fluorescence images (exposure \SI{500}{ms}) of single atoms at imaging detunings $\Delta/2\pi = -64.8, -23.7, -10.0, -3.2$ \SI{}{MHz} (lightening shades of blue). Near atomic resonance, single-atom fluorescence is easily distinguished from camera noise (gray), allowing us to detect an atom with $99.99\%$ fidelity.  For far-detuned probing, the photon-count distributions for zero- and one-atom images overlap ($\Delta/2\pi = -64.8$ MHz data shown together in inset, gray).  However, postselection (inset, blue) based on a later near-resonant image frame identifies single-atom images.
    (b) Detected photon counts vs $\Delta$ for an atom held in an optical tweezer of depth $k_B \times\SI{1.5(2)}{mK}$ (blue) and subject to increasing intensities of LW cavity field (lightening shades of red). Error bars indicate estimated single-shot uncertainty based on photon detection noise. Solid lines are fits to Eq.~\eqref{eq:brightness}. 
    The inset compares $\delta_\mathrm{ac}$ from LW light measured at $\Delta/2\pi = -37$ MHz [gray line in Fig. 2(b)] with a prediction that is based on the estimated cavity circulating intensity, with a correction factor of 0.7 applied to account for a reduction in the ac Stark shift measured at the LW antinode due to spatial averaging by the finite temperature atomic distribution.
    }
    \label{fig:Fig_2}
\end{figure}

The fluoresced light at a wavelength of \SI{780}{nm} is imaged using the same high-resolution objective used to produce the tweezer array.
In each experimental repetition, we image the same tweezer-trapped atoms in up to ten \SI{500}{ms} exposures, maintaining constant imaging-light intensity while varying the detuning $\Delta$ stepwise toward atomic resonance between image frames. 
Single atoms fluoresce more brightly in subsequent frames, as shown in the histograms in Fig.~\ref{fig:Fig_2}(a).
We postselect data in each image frame based on a photon-count threshold in a later frame to ensure the atom did not leave the trap during any earlier imaging exposure.
This postselection also allows us to determine the scattering rate for atoms probed with far-detuned light for which the fluorescence level is not far above the photodetection noise floor; see Fig.~\ref{fig:Fig_2}(a) inset.

The detected photon counts provide a measure of $\delta_\mathrm{ac}$, as demonstrated in Fig.~\ref{fig:Fig_2}(b).
We place a tweezer-trapped atom at the antinode of the LW cavity mode and detect atomic fluorescence at different imaging detunings $\Delta$ and linearly increasing LW cavity intensities. 
For each LW intensity, we fit the dataset of photon counts vs $\Delta$ to the prediction of Eq.~\eqref{eq:brightness} and observe good agreement with the model. 
As expected from the level structure of \Rb [see Fig.~\ref{fig:Fig_1}(c)], increasing LW light shifts the atomic resonance downward ($\delta_\mathrm{ac}<0$), bringing the effective imaging-light detuning $\Delta - \delta_\mathrm{ac}$ closer to resonance and increasing the atomic fluorescence.

\begin{figure}
    \centering
    \includegraphics{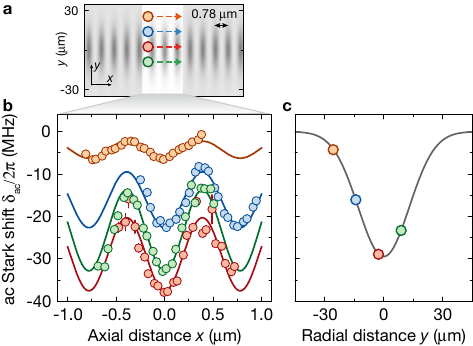} 
    \caption{Superresolution imaging of LW cavity field.
    (a) Several atoms trapped in tweezers of depth $k_B \times\SI{1.5(2)}{mK}$ are arrayed in a direction nearly perpendicular to the cavity axis, spaced to probe the cavity mode at several radial positions.  The calculated intensity profile of the cavity mode is shown in gray scale.  The array is translated in \SI{100}{nm} steps along the cavity axis (trajectory shown).  $\delta_\mathrm{ac}$ is measured for each tweezer at every position along this trajectory.
    (b) Axial dependence of ac Stark shift measurements for four tweezers whose radial positions are indicated in (c).  Modulation with a periodicity of the LW lattice ($d =$ \SI{780}{nm}) is clearly visible with contrasts between 0.30 and 0.47. Error bars indicate standard error on the mean of repeated measurements.
    (c) Radial profile of the LW TEM$_{00}$ mode, determined from axial average $\bar{\delta}_\mathrm{ac}$ of data in (b). Error bars are smaller than the data points. Solid line is a fit to a Gaussian with a waist of \SI{25.9(3)}{\micro m}.}
    \label{fig:Fig_3}
\end{figure}

For normal sensor operation, we determine the LW-light-induced $\delta_{\mathrm{ac}}$ from the photon counts detected at a single $\Delta$.
We convert the detected photons counts to $\delta_{\mathrm{ac}}$ using Eq.~\eqref{eq:brightness} with parameters, such as the ac Stark shift due to the tweezer light, calibrated by a fit to reference data taken in absence of LW light [blue data in Fig.~\ref{fig:Fig_2}(b)].
As shown in the Fig.\ \ref{fig:Fig_2}(b) inset, the  measured values of $\delta_\mathrm{ac}$ are consistent, to within systematic error, with our estimates of $\delta_{\mathrm{ac}}$ based on measurements of the LW power at the cavity output, thus demonstrating the accuracy of the sensor. The uncertainty in the estimate of $\delta_{\mathrm{ac}}$ is dominated by 20\% uncertainty in the transmissivity of the cavity out-coupling mirror~\footnote{To estimate the circulating power in the cavity, we account for $100$ ppm transmissivity of the out-coupling mirror and $10\%$ photodetection efficiency of transmitted light. The circulating power is converted to intensity via the measured mode waist [see Fig.~\ref{fig:Fig_3}(b)] and to $\delta_\mathrm{ac}$ by the scalar dynamical electrical polarizability at \SI{1560}{nm}.}.

The sensitivity of a single $\delta_\mathrm{ac}$ measurement is limited by photodetection noise, improving as $|\Delta - \delta_\mathrm{ac}|$ decreases and the atomic scattering rate increases, until the atom is lost from the tweezer trap owing to ineffective laser cooling under imaging light that is too close to atomic resonance. 
The best sensor performance that we demonstrate occurs at a minimum imaging detuning of $|\Delta-\delta_\mathrm{ac}|=2\pi\times\SI{30}{MHz}$. At this setting, pure shot noise on the photon number detected from a single atom would yield an ac Stark shift measurement sensitivity of $2\pi \times 250\, \text{kHz}/\sqrt{\text{Hz}}$.  In practice, read noise, background light, and additional noise due to atomic internal and motional dynamics increase the sensitivity to $2\pi\times500\, \text{kHz}/\sqrt{\text{Hz}}$. 

By scanning the positions of several tweezer-trapped atoms and performing a fluorescence measurement at each position, we obtain a scanning-probe image of the LW cavity mode, shown in Fig.~\ref{fig:Fig_3}.
We resolve both the coarse radial variation and also the fine-scale axial variation of the standing-wave Gaussian mode.
By averaging repeated axial scans of the cavity field, we identify and correct for a slow drift of the tweezer positions relative to the optical cavity of up to \SI{800}{nm}.

The contrast of the observed axial variation in $\delta_\mathrm{ac}$ provides a measure of the spatial resolution of our sensor.
The convolution of a full-contrast sinusoidal intensity pattern of period $d$ with a Gaussian point spread function of rms width $\sigma$ yields an expected contrast of $C = e^{-\pi^2 r^2 / 2 d^2}$ where $r=2\sigma$ is the resolution limit according to the Sparrow criterion.
The contrast achieved at various radial positions in Fig.~\ref{fig:Fig_3}(b) ranges between $0.30$ and $0.47$, which, with $d=$ \SI{780}{nm}, corresponds to $r$ between $380$ and \SI{300}{nm}. 
For comparison, the diffraction-limited Sparrow resolution of our $\mathrm{NA}=0.5$ microscope is 657 nm, and the fundamental free-space far-field diffraction limit for our imaging wavelength is 328 nm.
We achieve  superresolution with respect to both limits.
Decreasing contrast at higher cavity field intensity indicates that the temperature of the atomic sensor increases, perhaps owing to poorer laser cooling in the presence of large state- and position-dependent ac Stark shifts.
Using the axially averaged values of the ac Stark shift, $\bar{\delta}_\mathrm{ac}$, we reproduce the radial Gaussian profile of the cavity intensity, observing a beam waist of \SI{25.9(3)}{\micro\meter}, in agreement with our \textit{a priori} estimate of \SI{24.3}{\micro\meter} based on the cavity geometry~\footnote{The cavity mirrors have a radius of curvature of \SI{5}{mm} and the cavity length is \SI{9.4}{mm}.}.

\begin{figure}
    \includegraphics{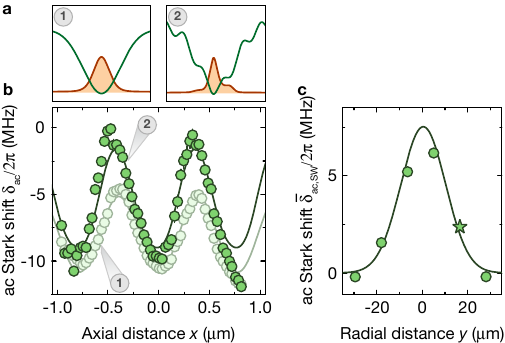} 
    \caption{Force sensing measurement of SW cavity lattice through distortion of LW measurement.
     (a) Optical trapping potential (green) and thermal atomic distribution (orange) due to the tweezer (left) and the sum of the tweezer and SW cavity field (right). The SW lattice narrows and displaces the atomic distribution.
     (b) Axial ac Stark shift due to the LW cavity field measured with a single atom trapped in a tweezer of depth $k_B \times\SI{0.34(3)}{mK}$ in the absence (presence) of SW light, shown in light (dark) green.  The axial variation at the LW periodicity gains an overall offset, increases in amplitude, and is shifted axially upon imposition of the SW potential. Error bars indicate standard error on the mean of repeated measurements.
     (c) Axially-averaged offset of the LW ac Stark shift due to the SW cavity field measured at different radial positions. Solid line is a fit to a Gaussian with a waist of \SI{20(3)}{\micro m}. Starred point indicates the tweezer used for (b). 
     }
\label{fig:Fig_4}
\end{figure}

Next, we explore the limits of the single-atom sensor by applying it to map an optical field generated by shorter-wavelength light, the standing-wave Fabry-P\'{e}rot cavity mode at a wavelength of \SI{781}{nm},  $2\pi\times\SI{400}{GHz}$ red-detuned from the atomic $D_2$ resonance. 
The shorter-wavelength light presents two coupled challenges to our sensing application.
First, at fixed resolution, shorter-wavelength axial intensity variation with periodicity $d =$ \SI{390.5}{nm} is measured with lower contrast, estimated at just $C= 0.05$ for $r = $ \SI{300}{nm}.  
For our measurement times and sensitivities, this low contrast allows only relatively large ac Stark shifts to be measured. 
Second, large ac Stark shifts modulated at short length scales lead to strong optical forces that displace the trapped atom within the optical tweezer, complicating the interpretation of the measurement.  
Further, unlike the LW light, the SW light produces ac Stark shifts on the ground and excited atomic states that are comparable in magnitude, exacerbating the deflection of the ground-state atomic sensor.

We overcome these obstacles by driving the Fabry-P\'{e}rot cavity simultaneously at both its SW and LW resonances and measuring the total transition ac Stark shift provided by both optical fields. 
In a simplified interpretation, we use the single atom now as a ``force sensor'': 
Forces and force gradients produced by the SW optical pattern displace and compress the space sampled by the trapped atom.
These changes to the atomic position distribution alter the observed axial variation of $\delta_\mathrm{ac}$ at the \SI{780}{nm} spatial period of the LW pattern, indirectly revealing the spatial structure of the SW mode; see Fig.~\ref{fig:Fig_4}.

We note three distinct features that arise due to the SW light.  First, the axial average of $\delta_\mathrm{ac}$ is shifted in the positive direction by an amount $\bar{\delta}_\mathrm{ac,SW}$ that is proportional to the circulating power of the SW light.
The radial variation of this average shift maps out the radial profile of the SW cavity mode; see Fig.~\ref{fig:Fig_4}(c).

Second, the SW light enhances the amplitude of the LW signal.
As shown in Fig.~\ref{fig:Fig_4}(a), for tweezers placed near antinodes of the SW cavity mode, the cavity-light potential adds to the tweezer confinement and reduces the size of the atomic distribution.  
If these antinodes overlap with the antinodes (nodes) of the LW standing wave, the narrowing of the atomic distribution increases (reduces) the magnitude of the ac Stark shift from the LW light.
Altogether, the contrast of the axial variation in the ac Stark shift measurement is thereby increased, consistent with our observation.

Third, the axial modulation pattern shifts along the axis in the presence of SW light.
This shift arises from an axial displacement of the SW cavity mode with respect to the LW mode, which is expected due to their noncommensurate wavelengths.
We perform a fit to the data accounting for all of these effects and determine the relative displacement between the LW and SW standing waves in the sensing region to be \SI{50(5)}{nm}. 
By measuring the amplitude and spatial displacement relative to the LW field, we have fully characterized the SW optical field. 


Directly mapping the spatial modes of our high-finesse optical cavity provides an excellent characterization of our system for future work in cavity QED. 
Complementing existing methods for measuring and controlling atom-cavity coupling~\cite{Purdy2010,Lee14,wu2021sitedependent,Fogliano2021}, this technique demonstrates our ability to position single tweezer-trapped atoms with subwavelength accuracy for cavity-mediated readout~\cite{Boozer2006,Khudaverdyan2009,Bochmann2010,Gehr2010}, feedback~\cite{Smith2002,Minev2019}, and entanglement~\cite{Welte2018,Vaidya2018,Davis2019,Samutpraphoot2020}.
High-resolution \textit{in situ} measurements of ac Stark shifts could be similarly useful in free-space quantum simulators and information processors that demand ever-better control of optical potentials~\cite{Zupancic16,Heinz2021}.

In the measurement demonstrated here, we use single-atom fluorescence on a strong allowed optical transition, measuring atomic energy shifts at the scale of the atomic linewidth (megahertz).
The measurement sensitivity could be improved by reducing photon detection noise to the shot-noise limit. 
One could extend this technique to detect weaker signals utilizing longer-lived atomic coherences, measuring local fields through their influence on spin coherences, for example, as is done in scanning probe microscopy using nitrogen-vacancy defects in diamond~\cite{Maletinsky2012}.
The spatial resolution of this measurement could be further improved by using a more tightly localized atomic probe, achievable with a deeper tweezer trap or by implementing demonstrated single-atom cooling techniques~\cite{Kaufman2012,Thompson2013,Yu2018}.
Implementing this measurement in a state-insensitive or ``magic wavelength'' tweezer trap would, by eliminating the ac Stark shift due to the tweezer, enable a calibration-free measurement of the target optical field~\cite{Arora2010,Aliyu2021}.

We thank A. Bohnett, S. Debnath, J. Ho, A. Lloyd, and R. Tsuchiyama for their assistance in the lab. 
We acknowledge support from  the AFOSR (Grant No. FA9550-19-1-0328), 
from ARO through the MURI program (Grant No. W911NF-20-1-0136), 
and from DARPA (Grant No. W911NF2010090).
E.D. and J.G. acknowledge support from the NSF Graduate Research Fellowship Program, 
and J.Z. acknowledges support from the Humboldt Foundation through a Feodor Lynen Fellowship.
 

\bibliography{cavity-microscope}

\clearpage

\end{document}